\documentclass[prl,twocolumn,pdflatex,showpacs,superscriptaddress]{revtex4}
\usepackage{amsmath, amsthm, amssymb,float}
\usepackage{amsfonts}
\usepackage{graphicx}
\usepackage{dcolumn}
\usepackage{bm}
\usepackage{textcomp}
\usepackage[normalem]{ulem}

\usepackage{color}

\usepackage{epsfig}
\usepackage{textcomp}
\usepackage{epstopdf}

\begin{document}

\title{Synergic Cherenkov-Compton Radiation }

\author{S.V. Bulanov}
\affiliation{Institute of Physics of the ASCR, ELI--Beamlines project, Na Slovance 2, 18221, Prague, Czech Republic}
\affiliation{National Institutes for Quantum and Radiological Science and Technology (QST),
Kansai Photon Science Institute, 8--1--7 Umemidai, Kizugawa, Kyoto 619--0215, Japan}
\affiliation{Prokhorov  General Physics Institute of the Russian Academy of Sciences, Vavilov Str.~38, Moscow 119991, Russia}
\author{P. Sasorov}
\affiliation{Institute of Physics of the ASCR, ELI--Beamlines project, Na Slovance 2, 18221, Prague, Czech Republic}
\affiliation{Keldysh Institute of Applied Mathematics, Moscow, 125047, Russia}
\author{S.\,S.\,Bulanov}
\affiliation{Lawrence Berkeley National Laboratory, Berkeley,CA 94720, United States of America}
\author{G.\,Korn}
\affiliation{Institute of Physics of the ASCR, ELI--Beamlines project, Na Slovance 2, 18221, Prague, Czech Republic}

\date{\today}

\begin{abstract}
An ultra-relativistic electron emits Cherenkov radiation in vacuum with an induced by strong electromagnetic wave refraction index larger than unity. During the interaction with this wave the electron also radiates photons via the Compton scattering. Synergic Cherenkov-Compton process can be observed by colliding laser accelerated electrons with a high intensity electromagnetic pulse. Extremely high energy photons cannot be emitted via the Cherenkov radiation because the vacuum refraction index tends to unity at these energies. Experiments on studying these phenomena will reveal the properties of vacuum predicted by nonlinear quantum electrodynamics.
\end{abstract}

\pacs{
{12.20.Ds}, {41.20.Jb}, {52.38.-r}, {53.35.Mw}, {52.38.r-}, {14.70.Bh} }
\keywords{ photon-photon scattering, QED vacuum polarization, Nonlinear waves}
\maketitle

\nopagebreak


 In quantum electrodynamics (QED) the photons interact with each other via virtual electron-positron pairs, which gives rise to a broad range of processes, 
in particular, related to vacuum polarization, vacuum birefringence \cite{BLP}, and other processes. The study of the photon-photon interactions is considered 
one of the most important applications of the high power laser facilities from the point of view of the fundamental science \cite{MKSC11, MTB06, MSh06, PMHK12, KBEP12, KKB19, PB19}. 
Among these interactions there are those that attract particular interest due to the fact that their description lies outside the framework of the perturbation theory 
\cite{KBEP12,   KH16, GKK18, DG00}. Arguably, the electron-positron pair production from vacuum by strong electromagnetic (EM) field is the most well known one. The characteristic field associated with his interaction is the QED critical field. It is also known as the Schwinger field \cite{BLP}. It is equal to $E_S=m_e^2c^3/e\hbar$, where $e$ and $m_e$ are the electron charge and mass, $c$ equal to the speed of light in vacuum, and $\hbar$  is the Planck constant. This field strength corresponds to the light intensity of the order of $10^{29}$ W/cm$^2$. Although the EM field intensity needed to observe pair production can be lowered by using special configurations of EM fields ~\cite{BMN10,GGH13} down to $10^{27}$ W/cm$^2$, such intensities are well beyond the reach of the next generation of laser facilities. However, the physics at the QED critical field level can be probed by high energy electron beams colliding with intense EM waves \cite{PMHK12}. For high enough electron (and photon) energy as well as for enough strong electromagnetic field such processes of electromagnetic interaction cannot be described within the framework of perturbation theory~\cite{Na69,R70} making them attractive for studies with high power lasers and accelerators of charged particles ~\cite{BIMR18, YMG18, PP18, Ild19}.

In what follows we consider the interaction of a high energy electron beam with 10 PW class lasers, focused to a one micron spot. 
In this case the laser intensity can reach $10^{24}$~W/cm$^2$, which corresponds to the laser EM field normalized amplitude of $a_0=10^3$. 
Here $a_0=eE_0/m_e\omega_0c$,  $E$  and $\omega_0$ are the laser electric field strength and frequency respectively.

The electron interacting with the field of the amplitude $a_0$ emits the photons with the energy   
$\hbar\omega_\gamma\approx \hbar\omega_0 a_0\gamma_e^2$, where $\gamma_e$ is the elecron Lorentz factor. 
Estimating the electron quiver energy as $\gamma_e\approx a_0$,  we find that the photon energy for 
$a_0\approx 10^3$  and $\hbar\omega_0=1$~eV is in the $\gamma$-ray range. 
As we see, an electron interacting with the electromagnetic wave emits high order harmonics. 
The maximum harmonic number could be approximately equal to $a_0^3$. In quantum physics, 
this corresponds to absorption of $N_{ph}=a_0^3$ laser photons by an electron in order to emit one high energy photon. 
The quantum effects come into play when $\hbar\omega_\gamma\approx m_e c^2 \gamma_e$  and  $a_0\gg 1$, 
and the photon emission occurs in the multi-photon Compton scattering process (see Ref.~\cite{DelC19}   and review article \cite{EKK09} and references cited therein).

The Lorentz invariant parameters,   $\chi_e=\sqrt{p_\mu F^{\mu\nu}F_{\nu\rho}p^\rho}/m_ecE_S$ and  
$\chi_\gamma=\hbar\sqrt{k_\mu F^{\mu\nu}F_{\nu\rho}k^\rho}/m_ecE_S$, characterize the QED processes 
for electrons and photons interacting with the electromagnetic field. Here $p_\mu$  is the electron 4-momentum,  
$k_\mu$ is the photon 4-wave vector, $F_{\mu\nu}=\partial_\mu A_\nu-\partial_\nu A_mu$  is the electromagnetic 
field tensor, and  $A_\mu$ is the electromagnetic 4-potential. 
If the parameter  $\chi_e$  is equal to unity, in the electron rest frame, the electric field is equal 
to the critical QED field $E_S$. 

Strong electromagnetic wave induces the vacuum polarization. 
As a result, in the long wavelength limit, the QED vacuum behaves as 
a medium with a refraction index larger than unity~\cite{GKK18, DG00, Na69, R70}, 
i. e. the speed of propagation of the interacting electromagnetic waves is below speed of light in vacuum. One of the 
consequences of this fact is a possibility of the Cherenkov radiation of the high-energy electrons traversing the 
electromagnetic field~\cite{Na69, R70, D02, AJM18, Schw76, B77,  GinTs79}. In Refs. \cite{D02, AJM18} the 
electromagnetic field was considered to be generated by exteremely high power lasers.

Below we analyze the properties of Synergic Cherenkov-Compton Radiation and Scattering (SCCRS) and discuss a way for observing this phenomenon with extreme high power lasers by colliding laser-accelerated electrons with high intensity electromagnetic pulse.

When ultra-relativistic electron collides with the EM wave  it undergoes radiation losses which can prevent the electron 
from reaching the the high intensity EM field region. To describe the one-dimensional relativistic electron dynamics in 
the EM field, following to approach formulated in Ref. \cite{bulanov.pra.2013}, we use a system of differential equations for the distribution functions of electrons, positrons and photons with the  
%
%
%
the differential probabilities 
of a photon emission by an electron/positron, and a photon decay into electron-positron pair (for details see Ref. \cite{bulanov.pra.2013}). It is emphasized here that they are the functions of initial electron/positron/photon energy and the instantaneous (at time $t$) value of the EM field. Here we assume that the electron, positron, and photon dynamics is dominated by the longitudinal motion. 

For a 10~PW laser focused to a one-lambda spot the normalized amplitude approximately equals to $a_0=10^3$. Here we assume that the characteristic size of the high intensity region is equal to one lambda. As the energy of the laser wake field accelerated (LWFA) electrons according to the LWFA scaling~\cite{ESL09} is ${\cal E}_e=10\, (P/1~\mbox{PW})$~GeV, then for a 10~PW laser it can reach 100~GeV, i.e. $\gamma_{in}=2\times10^5$. Solving a system of equation 
for the electron, positron and photon distribution functions for these initial parameters and assuming the electron beam is monoenergetic before the interaction, we obtain a broad spectrum of electrons after the interaction. The electron beam lost approximately half of its energy, however 37\% of electrons has energy above 40 GeV. The situation at the maximum of the EM field is slightly different. Approximately 7\% of the initial electrons reach the maximum of the EM field without emitting a photon. Moreover, around 60\% of electrons have energy higher than 40 GeV. The solution shows that a significant number of electrons with very high energies can reach the maximum of the EM field of the tightly focused laser pulse.


Here, we consider the kinematics of an inverse multiphoton scattering process when an ultra-relativistic electron 
collides with an electromagnetic wave. Before and after the collision the electron has the momentum  $\pmb{p}_0$  and 
energy
$m_ec^2\gamma_{0}=\sqrt{\pmb{p}_0^2c^2+m_e^2c^4}$ and the momentum   $\pmb{p}$ and energy 
$m_ec^2\gamma_{e}=\sqrt{\pmb{p}^2c^2+m_e^2c^4}$, respectively. The scattered photons have the frequency   
$\omega_0$ before collision and $\omega_\gamma$  after it. By using the energy and momentum conservation in the 
electron-photon system, we can find the scattering photon frequency dependence on the electron energy, the wave 
amplitude, and the scattering angle. As the electron interacts with $s$ photons from the wave and emits high energy photon, the energy and momenta of all particles are connected by the conservation laws, which yield: 
\begin{equation}\label{E090}
m_ec^2\gamma_{0}+s\hbar\omega_0=m_ec^2\gamma_e+\hbar\omega_\gamma\, 
\end{equation}
and
\begin{equation}\label{E100}
\pmb{p}_0+s\,\hbar\pmb{k}_0=\pmb{p}+\hbar\pmb{k}_\gamma\, ,
\end{equation}
where $\hbar \pmb{k}_0$ is the EM wave photon momentum and $\hbar \pmb{k}_\gamma$  is the emitted photon momentum.  

We consider the case when photon-electron interaction occurs in the   
$(x,y)$ plane, i.e.,  $\pmb{p}_0=p_{\parallel,0}\pmb{e}_{x,0}+p_{\perp,0}\pmb{e}_y$ and  
$\pmb{p}=p_\parallel\pmb{e}_{x}+p_{\perp}\pmb{e}_y$. For the perpendicular component of the momentum 
$p_{\perp,0}$  we may assume that it is equal to the quiver electron momentum in the electromagnetic field of the 
amplitude $a_0$ , i.e. the electromagnetic wave is linearly polarized with the electric field along the $y$  axis.

In the case of classical electrodynamics described by the Maxwell equations, 
the electromagnetic wave frequency and wave vector are related to each other as $\omega^2=\pmb{k}^2c^2$ . 
In quantum vacuum the polarization effects result in the dispersion equation ~\cite{Na69,R70}

\begin{equation}\label{E120}
\omega^2-\pmb{k}^2c^2-\mu_\pm^2\left(\chi_e\right)c^2\hbar^{-2}=0\, .
\end{equation}
Here $\mu_\pm$ is the invariant photon mass~\cite{Na69,R70}.
The signs $\pm$  in a subscript of $\mu_{\pm}$ correspond to the parallel and perpendicular polarizations of the 
colliding electromagnetic waves. The invariant mass depends on the photon 
frequency (it is the photon energy expressed in terms of the quantum parameter $\chi_\gamma$). Its square is given  
asymptotically by
\begin{equation}
\label{E130}
\mu_\pm^2=-\alpha m_e^2\left[
\frac{11\mp 3}{90\pi}\chi_\gamma^2+i\sqrt{\frac{3}{2}}\frac{3\mp 1}{16}\chi_{\gamma}
\exp\left(-\frac{8}{3\chi_\gamma}\right)\right]
\end{equation}
for $\chi_\gamma\ll 1$ and
\begin{equation}
\label{E131}
\mu_\pm^2=-\alpha m_e^2\left[\frac{5\mp 1}{28\pi^2}\sqrt{3}\Gamma^4\left(\frac{2}{3}\right)\left(1-i\sqrt{3}\right)(3\chi_\gamma)^{2/3}\right]
\end{equation}
for $\chi_\gamma\gg 1$.
Here $\Gamma(x)$  is the Euler gamma function. The imaginary part of $\mu_\pm^2$  gives a probability 
of the electron-positron pair creation. As noted in Ref.~\cite{R70}, for large   $\chi_\gamma$, at 
$\alpha\chi_\gamma^{2/3}\sim 1$ , the photon mass becomes of the order of the electron mass. In this limit, the 
perturbation theory becomes inapplicable.

Eqs.~(\ref{E120}-\ref{E131}) yield that, in the limit $\chi_\gamma\ll 1$ , 
a difference between the refraction index value and unity, $n_{\pm} - 1=\Delta n_{\pm}$. It is
\begin{equation}\label{E140}
\Delta n_\pm=\alpha\frac{11\mp 3}{45\pi}\left(\frac{E}{E_S}\right)^2\, .
\end{equation}
From this expression it follows that in the low photon energy limit,  $\chi_\gamma\ll1$ the normalized phase velocity
(and equal to it group velocity), $\beta_{\pm}=v_{\pm}/c$,  of the linearly polarized counter-propagating electromagnetic 
wave with the electric field parallel or perpendicular to the $y$  axis equals 
$\beta_{ph,\pm}=1-\varepsilon_\pm(E_0/E_S)^2$, where the coefficients  
$\varepsilon_\pm$ are $\varepsilon_\pm=\alpha\, (11\mp3)/45\pi\approx 10^{-4}$.

Now for the sake of brevity we assume that the transverse component of electron momentum before interaction 
vanishes, i. e.  $\pmb{p}_0=p_{\parallel,0}\pmb{e}_x$. From Eqs.~(\ref{E120}-\ref{E140}) follows that, as a 
consequence of the vacuum polarization, the relationship between wave number and frequency of the electromagnetic \
wave corresponding to the wave propagating in a medium with the refraction index not equal to unity can be written as
\begin{equation}\label{E150}
\pmb{k}=\frac{\pmb{k}}{|\pmb{k}|c}n_\pm\, \omega\, ,
\end{equation}
with $\pmb{k}=|\pmb{k}|\left(\pmb{e}_x\cos\theta+\pmb{e}_y\sin\theta\right)$, where $\theta$  is the angle between the 
scattered photon wave number and the $x$  axis.  Using this relationship to solve Eqs.~(\ref{E090}) and~(\ref{E100}), 
we obtain for the energy of scattered photon
\begin{equation}
\label{E160}
\hbar\omega_\gamma=g \pm \sqrt{g^2+2s\,\hbar\omega_0
\left(
\frac{m_e c^2 \gamma_0+p_{\parallel,0}c
}{n_\pm^2-1}
\right)},
\end{equation}
where
\begin{equation}
\label{eq:fg}
 g=\frac{\left(p_{\parallel,0}c - s\,\hbar\omega_0\right) n_\pm \cos\theta-
m_e c^2 \gamma_0-s\,\hbar\omega_0}{n_\pm^2-1}.
\end{equation}
Only positive solution for the photon frequency $\omega_\gamma$ is relevant to our problem. 

The equation (\ref{E160}) describes kinematics of the SCCRS process in QED vacuum, whose optical 
properties are  modified by the strong electromagnetic wave.

In the case, when the modification of the vacuum refraction index is weak enough, i.e. when $0\leq n_\pm^2-1\ll 1$,  
 the dependence on the parameters of the emitted photon frequency  $\omega_\gamma$ given by Eq, ~(\ref{E160})  
 can be presented as a combination of two modes with a continous transition between them. When the function $g$ 
 given by Eq. (\ref{eq:fg}) is positive  
and $s\, \hbar\omega_0\ll m_ec^2/ \gamma_e$ the photon energy can be found to be  
\begin{equation}
\label{E170}
\hbar\omega_{Ch}\approx2g
+s\,\hbar\omega_0\left[\frac{
 m_ec^2\gamma_0+p_{\parallel,0}c}{g (n_\pm^2-1)}\right]\,. 
\end{equation}
This expression corresponds to the Cherenkov radiation with a notation $\omega_{Ch}$ used 
for the frequency of the photon emitted in this regime.

In the opposite limit, when the function $g$ 
 given by Eq. (\ref{eq:fg}) is negative, in the limit  
$s\, \hbar\omega_0\ll  m_ec^2/\gamma_e$, Eq.~(\ref{E160}) can be rewritten as
\begin{equation}
\label{E180}
\hbar\omega_{C}\approx
-\frac{s\,\hbar\omega_0(m_ec^2\gamma_0+p_{\parallel,0}c)}
{m_e c^2 \gamma_0+s\,\hbar\omega_0-\left(p_{\parallel,0}c - s\,\hbar\omega_0\right) n_\pm \cos\theta}\, ,
\end{equation}
corresponding to the Compton scattering mode with the photon frequency equal to $\omega_{C}$. In the limit $s\ll s_m$, where
\begin{equation}
\label{eq:Nphot}
s_m= \frac{m_ec^2}{4  \hbar \omega_0\, \gamma_0}
\end{equation}
with $\gamma_0=\sqrt{1+p_{\parallel,0}^2/m_e^2c^2}$. The photon energy  equals  
$\hbar\omega_{C}\approx 4 s\,\hbar\omega_0 m_ec^2\gamma_0^2$. At $s\geq s_m$ we have 
$\hbar\omega_{C}\approx  m_ec^2\gamma_0$.

The condition  $s\leq s_m$ shows when 
the recoil effects due to the photon emission become dominant. Eq. (\ref{eq:Nphot})   gives the photon number 
absorbed by the electron with the energy $m_ec^2 \gamma_0$ from the electromagnetic wave for radiating one high energy photon. 

Assumption $0\leq n_\pm^2-1\ll 1$ in Eq.~(\ref{E170}) yields a condition of Cherenkov radiation,
\begin{equation}
\label{E190}
 n_\pm>\frac{\sqrt{m_e^2c^2+p_{\parallel,0}}}{p_{\parallel,0}}\, ,
\end{equation}
imposing the requirement on the electron energy~\cite{GinTs79, D02}. We note that this condition is written 
assuming that 
\begin{equation}
\label{eq:som}
s\, \hbar\omega_0\ll \frac{\sqrt{m_e^2c^4+p_{\parallel,0}^2c^2}-p_{\parallel,0}c n_\pm \cos\theta }
{1+n_\pm \cos\theta}.
\end{equation}
The electron energy should be large enough to have 
\begin{equation}
\label{E210}
\gamma_0>\gamma_{Ch}=\frac{1}{\sqrt{2\Delta n_\pm}}=\sqrt{\frac{45 \pi E_S^2}{\alpha (11\mp3)E_0^2}}\approx 
30\sqrt{\frac{I_S}{I_0}}\, .
\end{equation}
Here  the laser intensity $I_0=cE_0^2/4\pi$, which in the focus region of 10~PW laser approximately is equal to  
$10^{24}$~W/cm$^2$ and  $I_S=cE_S^2/4\pi\approx 10^{29}$~W/cm$^2$, i. e. the Cherenkov radiation threshold is 
exceeded for the electron energy above 10~GeV. 


\begin{figure} [ht]
\includegraphics[width=0.49\textwidth,clip=]{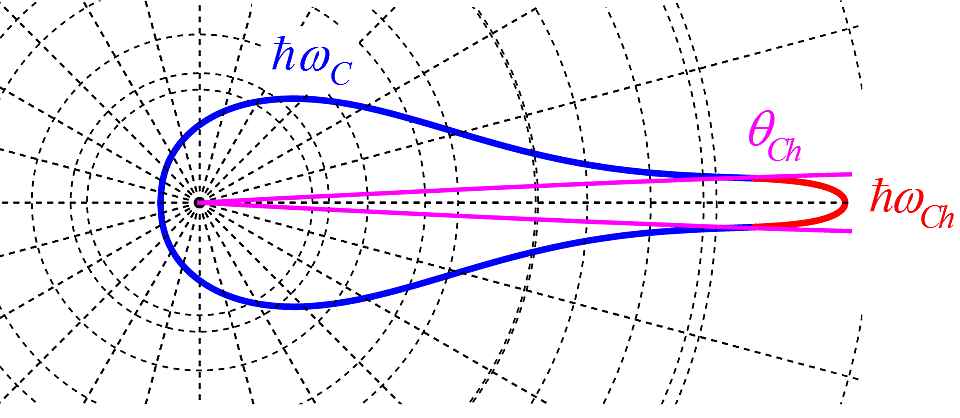}
\caption{Angle distribution of the energy logarithm for photon radiated by the SCCRS mechanism. 
Blue color used for Compton scattering, $\hbar \omega_{C}(\theta)$, and red color for the Cherenkov radiation, $\hbar \omega_{Ch}(\theta)$. Magenta lines $\theta=\pm \theta_{Ch}$ show the Cherenkov cone.
}
\label{fig1}
\end{figure}

We can see from Eqs.~(\ref{E170}) and~(\ref{E180}) that photons emitted via Cherenkov and Compton mechanisms 
have different angular  distribution. The photons emitted via the Cherenkov radiation process are confined within the 
Cherenkov cone with the angle  $\theta_{Ch}=2\sqrt{\varepsilon_\pm\, I_0/I_S}$. In the focus of 10~PW laser it is 
approximately equal to $ 2\times 10^{-5}$. The Compton scattered photons are within the cone with the angle   $
\theta_C\approx 1/\gamma_0$. Although characteristic angle values for these process are of the same order, the 
dependence of the 
photon  energy on the angle is different as it is seen from Fig. \ref{fig1}, where we show the
angle distribution of the photon energy logarithm for Cherenkov radiation and for Compton scattering. 
Photons of substantially low energy emitted via the Compton scattering are localized within the relatively 
wide cone (blue curve). The more narrow beam of the Cherenkov radiation photons with higher energy is shown with the red color. The Cherenkow cone $\theta_{Ch}$ is shown by the lines $\theta=\pm \theta_{Ch}$.

According to the Cherenkov radiation theory~\cite{TF37} the rate of the energy loss due to the Cherenkov radiation friction force along the electron trajectory is
$$
\frac{d{\cal E}_e}{dx}=-\frac{e^2}{c^2}\int\limits_{v_en/c>1}\left(1-\frac{c}{v_en_\pm}\right)\,
\omega d\omega
$$
\begin{equation}\label{E220}
\approx-\frac{e^2}{\lambdabar_C^2}\, \varepsilon_\pm\, \left(\frac{E_0}{E_s}\right)^2\, .
\end{equation}
Integration is done over the region where $v_en/c>1$ with $v_e=c p_{||,0}/\sqrt{m_e^2c^2+p_{||,0}^2}$ is the electron velocity. 

The formation length~\cite{BaKa05} in the case of Cherenkov radiation is given by
\begin{equation}\label{E230}
\ell_{Ch}\approx\gamma_e\gamma_{Ch}\lambdabar_{Ch}\approx\lambdabar_C\gamma_e\, .
\end{equation}
 Here  $\lambdabar_C=\hbar/m_e c \approx 3.8\times 10^{-11}$~cm is the Compton scattering wavelength~\cite{BLP},   
 $\lambdabar_{Ch}\approx\lambdabar_C/\gamma_{Ch}$ is the wavelength of radiation emitted via the Cherenkov 
 radiation mechanism, and $\gamma_{Ch}$   is the threshold energy determined by expression (\ref{E210}). For 
 10~PW laser radiation the electron threshold energy is approximately equal to 10~GeV. According to Eq. (\ref{E230}) 
 the radiation formation length is   $\ell_{Ch}\approx5\times10^{-5}$~cm, i. e. it is comparable 
 in magnitude with  the laser wavelength. Traversing the laser focus region the electron emits 0.2 photons. 
 Assuming the electric charge of the LWFA electron bunch of 100 pC we obtain $10^4$  photons (the efficiency of the 
 electron energy conversion to the Cherenkov radiation is  $\approx 10^{-4}$). Photons produced in the Compton 
 scattering have approximately the same frequency. Interaction of the Compton scattering photons with the laser field 
 will result in the Breit-Wheeler electron-positron pair plasma generation \cite{BBK99, BellKirk}.

 As one may see from the expression for the photon invariant mass~(\ref{E130}) at the high photon energy
end, when the parameter $\chi_\gamma$ becomes larger than unity, the vacuum polarization effects weaken.
In this limit the Cherenkov radiation does not occur. As a result the photons with the energy above
$\hbar\omega_\gamma\approx m_e c^2 E_S/E_0$ are not present in the high frequency spectrum of the radiation.
 For 10~PW
laser parameters this energy is approximately equal to 100~MeV.

{
Considering kinematics of the process $e\to e\gamma$ in a strong electromagnetic field with taken into account  the 
radiation correction of the ``photon mass'', i. e. $\mu_\pm^2\ne 0$ given by Eq.~(\ref{E120}), we  neglect 
the electron mass radiation correction. According to Ref.~\cite{R70} the radiation correction to the electron mass 
scales as $\Delta m_e^2\propto \alpha \chi_e^{2/3}$ in the limit $\chi_e^{2/3}\gg 1$. The Cherenkov radiation condition 
(\ref{E210}) can be written in the form $c-v_\gamma>c-v_e$, where $v_\gamma$ and $v_e$ are velocity of scattered 
photon and the electron velocity. In the right hand side of this inequality, the radiation correction of the electron mass 
changes the electron velocity  approximately on 
$\alpha m_e^2\chi_e^{2/3}$. Thus, the radiation correction of electron mass for the Cherenkov radiation process 
would become important when $\chi_e\gtrsim \alpha^{-3/2}$, whereas the Cherenkov radiation condition requires 
$\chi_e\gtrsim \alpha^{-1/2}$. Such relationship indicates that the first experiments on 
Sinergic Cherenkov-Compton process revealing modification of virtual electron-positron 
sea in a strong electromagnetic field are expected to be far from the conditions, when the radiation correction of 
the electron mass would become important because the electrons and the scattered Cherenkov photons have quite different energies.
}

In conclusion, a scheme of the experiments on the laser accelerated electron interaction with focused EM wave aimed
at studying such fundamental physics processes as the radiation friction effects, electron-positron pair creation, and
vacuum polarization has been considered theoretically in Refs.~\cite{BBK99, NIMA12, AGRT12}. Its principle 
setup was realized in the experiments whose results are presented in Refs.~\cite{BBK99, COLE18, PODER18}.
Here we pay attention to the fact that the electron undergoing multi-photon Compton scattering also emits the
Cherenkov radiation in the QED vacuum, where a strong electromagnetic field induces a refraction index larger than
unity, thus entering the regime of Synergic Cherenkov-Compton Radiation-Scattering. In the range of the parameters
where the Cherenkov and Compton modes can be distinguished the angle and energy distributions for the gamma
photons emitted by these two mechanisms are different. With extreme high power lasers (they are of the type of
lasers build within the ELI project) the synergic Cherenkov-Compton process can be observed by colliding laser
accelerated electrons with a high intensity electromagnetic pulse. At extremely high photon energy end, when the
parameter $\chi_{\gamma}$ becomes larger than unity, as shown in Refs. \cite{Na69, R70}, 
the vacuum refraction index tends to unity thus
quenching the Cherenkov radiation. Observation of these phenomena can shed light on the properties of nonlinear
QED vacuum, allowing us to reveal the physical processes on a way towards the limit when
$\alpha \chi_{\gamma}^{2/3}$, i.e. when the electromagnetic field interaction with charged particles develops 
according to unperturbative regime scenario.

In the experiment under dicussion, a single laser pulse can be used within the framework of the all-optical scheme, 
to accelerate ultra-relativistic electrons in the laser-plasma interaction and then to be focused in the $\lambda^3$ region to achieve extreme high field amplitude providing conditions required for colliding the electrons with the 
electromagnetic field. Similar scheme is proposed in Ref. \cite{ALLOPT} for a Compton source based on the combination of a laser-plasma accelerator and
a plasma mirror. Using this scheme, the Synergic Cherenkov-Compton Radiation-Scattering in the interaction of
100 GeV LWFA accelerated by 10 PW laser electrons with the electromagnetic field of $10^{24}$W/cm$^2$ will 
reveal the vacuum properties predicted by nonlinear QED theory.

\begin{acknowledgments}

The work is supported by the project High Field Initiative (CZ.02.1.01/0.0/0.0/15\_003/0000449)from the European Regional Development Fund. SSB acknowledges support from the Office of Science of the US DOE under Contract No. DE-AC02-05CH11231.

\end{acknowledgments}


\end{document}